# Towards rapid extracellular vesicles colorimetric detection using optofluidics-enhanced color-changing optical metasurface

Chuchuan Hong[1,2], Ikjun Hong[1,2], Sen Yang[1,2], Justus C. Ndukaife[1,2,3,*]

1. Department of Electrical and Computer Engineering, Vanderbilt University, Nashville, TN, USA
2. Vanderbilt Institute of Nanoscale Science and Engineering, Vanderbilt University, Nashville, TN, USA
3. Department of Mechanical Engineering, Vanderbilt University, Nashville, TN, USA

Abstract:

Efficient transportation and delivery of analytes to the surface of optical sensors are crucial for overcoming limitations in diffusion-limited transport and analyte sensing. In this study, we propose a novel approach that combines metasurface optics with optofluidics-enabled active transport of extracellular vesicles (EVs). By leveraging this combination, we show that we can rapidly capture EVs and detect their adsorption through a color change generated by a specially designed optical metasurface that produces structural colors. Our results demonstrate that the integration of optofluidics and metasurface optics enables robust colorimetric read-out for EV concentrations as low as $10^7$ EVs/ml, achieved within a short incubation time of two minutes, while using a CCD camera or naked eye for the read-out. This approach offers the potential for rapid sensing without the need for spectrometers and provides a short response time. Our findings suggest that the synergy between optofluidics and metasurface platforms can enhance the detection efficiency of low concentration bioparticle samples by overcoming the diffusion limits.

**Table of Contents Figure:**

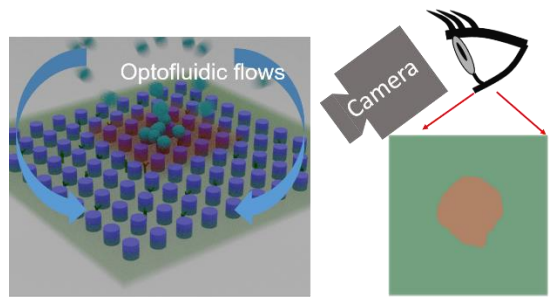

**INTRODUCTION:**

A metasurface is a planar nanostructured surface that can control the properties of light, such as phase, amplitude, or polarizations by designing the geometry, the arrangement, or the material of meta-atoms. One notable application of optical metasurface is to generate structural coloration in miniaturized systems[1–3], where those supporting structural coloration are referred to as "color metasurfaces" in this paper. Structural coloration studies the color generation from the interaction of incident light with wavelength/subwavelength scale structures. Dynamic structural coloration leverages the tunable response of the optical elements to the change of environmental optical properties, such as the refractive index of the surrounding media. This ability to produce real-time optical response forms the foundation for colorimetric sensing, which has been widely used in the detection of humidity[4,5], pH[6], chemical compounds[7,8], or biological analytes[9–11]. One essence of using color metasurfaces is that they can be readily interrogated by using the human eye or a digital camera to read out the signal, whereas high-Q metasurfaces requires sophisticated spectrometers or advanced laser systems for high sensitivity[12–15]. The sensitivity of the metasurface-based colorimetric sensor can be optimized by a well-designed optical response at visible wavelength.

Apart from high sensitivity detection, there is also a need to rapidly deliver analytes to the sensing surface without having to rely on slow Brownian diffusion. Most state-of-the-art biosensors face a general bottleneck due to Brownian diffusion limitation[16–19], which fundamentally hampers the overall performance of a biosensor. The process of transporting analyte onto the detector via stochastic Brownian motion results in an unpredictable and long duration for incubation, ranging from several hours to multiple days, particularly in solutions with low analyte concentrations[16–20]. A long incubation time makes the sensing in low analyte concentration solution impractical, consequently hindering the utilization of a sensor's detection capacity. Hence, there is a need to develop an active method to rapidly transport analytes and speed up the incubation process.

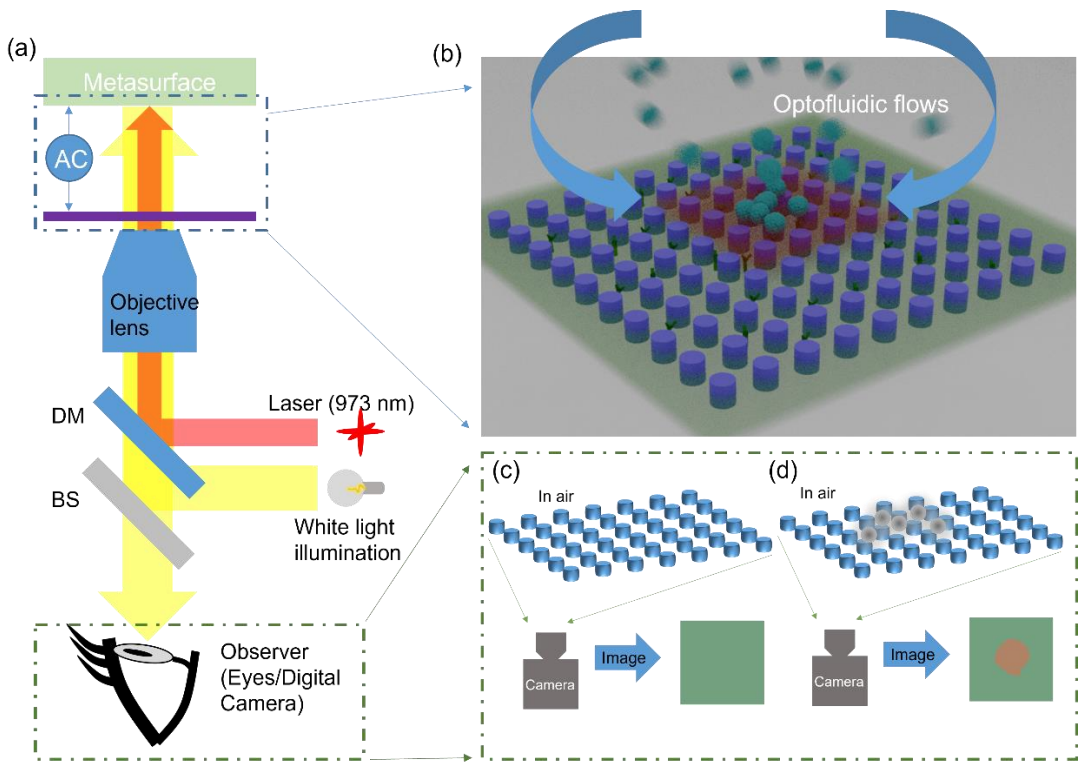

Figure 1: (a) Experiment setup for fast incubation and color metasurface imaging. (b) Schematic illustration of the metasurface's color change triggered by EVs (extracellular vesicles) binding to the surface. Using optofluidic body flow to rapidly concentrate particles and bring them down to the nanostructured substrate, the local color would change due to the refractive index contrast induced by the EV capture. (c) Before EV concentration on the sensing surface, and (d) after EV concentration on the sensing surface. The color changes over the region with the captured EVs. The optimized metasurface enables a highly perceivable color change imaged by the CCD camera.

In this paper, we report a silicon-based color metasurface with high-contrast color responses that can be used for label-free and high-speed colorimetric sensing when combined with optofluidics. To address the diffusion limitations and promote active transport to the sensing sites, we have incorporated optofluidic flows to transport analytes[21–23]. The setup we used for this demonstration is schematically illustrated in Fig. 1a. We employ a 973 nm laser to generate local heating atop the metasurface, which in turn creates electrothermoplasmonic (ETP) flows to rapidly transport particles toward the metasurface as we apply an a.c. electric field across the microfluidic channel[16,23–25]. This transport process is schematically illustrated in Fig. 1b. The maximum temperature rise should be well-managed so that it is sufficient for ETP generation but does no harm to biological particles like extracellular vesicles. Also, for the purpose of generating local heating, a thin layer of gold (8 nm) was deposited on the metasurface with a 3 nm chromium adhesion layer. ETP flows allow for rapidly transporting particles within a range of hundreds of microns and concentrating them on the illuminated surface. The signal read-out is accomplished by a digital camera (Nikon DS-Fi1). As shown in Fig. 1c and d, the existence of captured particles changes the local refractive index on the color metasurface and induces perceivable color change locally.

**EXPERIMENTS AND DISCUSSIONS:**

The rest of the manuscript is described as follows. First, we discuss the design of the color metasurface to generate structural colorations as the refractive index of the surrounding is tuned. Next, we discuss the optofluidic control enabled by ETP flow for rapid enrichment of EVs on the metasurface. Multiphysics modeling was used to simulate the temperature rise across the metasurface containing the chromium and gold coating layers, which contribute to the overall optical absorption and heat generation, as well to model the ETP flow velocity. Finally, to test the feasibility of achieving read-out of EVs adsorption, we concentrated EVs, which are negatively charged, on the metasurface after functionalizing with cysteamine, which is positively charged. We note that for future applications involving the capture and quantification of EVs for diagnosis and treatment, the cysteamine layer may be replaced by a specific antibody targeting EV surface protein markers.

To achieve the proposed dynamic structural colorations, we utilized the strong dispersion and resonances of the array of silicon nanopillars at visible wavelengths[3]. The multipolar resonances of high-index materials have been shown to efficiently modulate light scattering[26–28] and create chromatic responses[1,3]. Of note, the ability to produce this vivid chromatic response is not a universal feature of arbitrary metasurfaces. In our optimized color metasurface, the silicon nanopillars have dimensions of 72 nm in height, 90 nm in radius and 280 nm in lattice constant. The silicon nanopillars sit on a glass substrate. This optimized metasurface harnesses the interplay between the symmetric and antisymmetric modes to achieve sensitive chromatic tunability.

To better interpret the interaction between symmetric and antisymmetric modes in the color metasurface, we employed a multipolar analysis and reconstructed the reflection spectrum based on the extracted multipolar information[29,30]. We found that two dominant resonant modes exist in the visible wavelength range of interest. Their electric field distribution is illustrated in Fig. 2a and b, with Fig. 2a illustrating the symmetric mode profile and Fig. 2b showing the antisymmetric distribution. Additional information about the calculation details of the multipolar analysis is provided in Supplementary Section 3. Based on the symmetry of the multipolar radiation pattern with respect to the metasurface plane, the symmetric mode is primarily contributed by electric dipole ($ED_x$) and magnetic quadrupole ($MQ_{yz}$), while the antisymmetric mode arises from the interference of magnetic dipole ($MD_y$) and electric quadrupole ($EQ_{xz}$). Both the symmetric mode and antisymmetric mode undergo spectral shifts and change in strength as the environmental refractive index increases. As shown in Fig. 2c and d, when the refractive index of the environment increases from air (n = 1) to EV (n = 1.38), the dominating symmetric (electric dipole) response is enhanced and redshifted, while the antisymmetric (magnetic dipole) is suppressed. This results in a redshifted reflection spectrum with an enhanced value on the red wavelength end, which leads to a color change from green to red on the metasurface when the index of superstrate increases from air to say water.

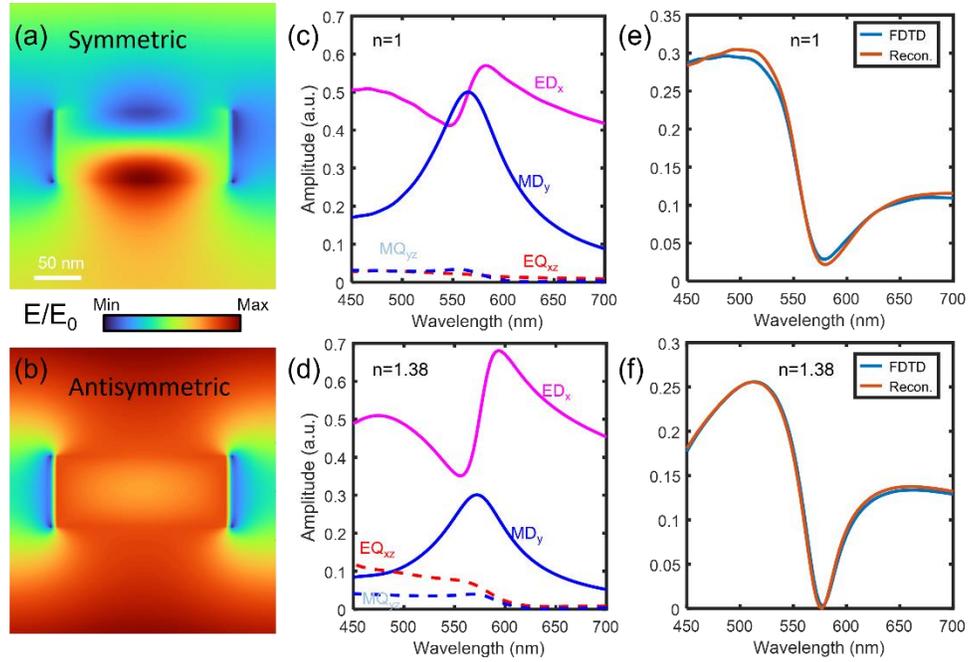

Figure 2. Electric field distributions of the (a) symmetric and (b) antisymmetric modes at visible wavelength, respectively. (c) Amplitudes of the scattered plane waves originated from various multipolar components in air (n = 1) and (d) in EVs (n = 1.38) respectively. (e) Reconstructed reflection spectra from multipole decomposition analysis in (c) and (f) reconstructed from (d). They show good agreements with simulated results.

As a control group, we also fabricated another metasurface with an altered lattice constant and pillar radius. The measured spectra of the control group are included in Fig. S2a. Although the control group shows a variance of reflection signal beyond 700 nm wavelength, the calculated and observed color changes both demonstrated trivial color shift. The optical images of the control group metasurface under the same imaging condition are also provided in Fig. S2b and c.

The SEM micrograph in Fig. 3a displays the fabricated color metasurface comprised of amorphous silicon pillars on the glass substrate. The measured and simulated reflection spectra (in Fig. 3b) show good agreement for both water and air. Notably, an increase of reflectance was observed in water at wavelengths greater than 600 nm, suggesting a change in color when placing the metasurface in water. We used a halogen lamp to illuminate the metasurface and observed its color under a microscope using an objective lens with a magnification of 40X and a numerical aperture of 0.75. The color appears green initially in air, as illustrated in Fig. 3c, and changes into red upon submersion in water as illustrated in Fig. 3d. This perceived color change is highly consistent with the predicted color change (from Fig. 3e(i) to 3e(ii)) based on the measured spectra presented in the color space of Fig. 3e, where the 1931 CIE (x,y) chromaticity coordinates are calculated from the measured spectrum[11,31]. The black circles marked (i) and (ii) indicate the colors reflected by the metasurface in air and water, respectively, and the corresponding colors are also highlighted in the insets. The calculated colors of the control group in air and water are also presented in the insets of (iii) and (iv), respectively, as marked by the white circles in the color space plot.

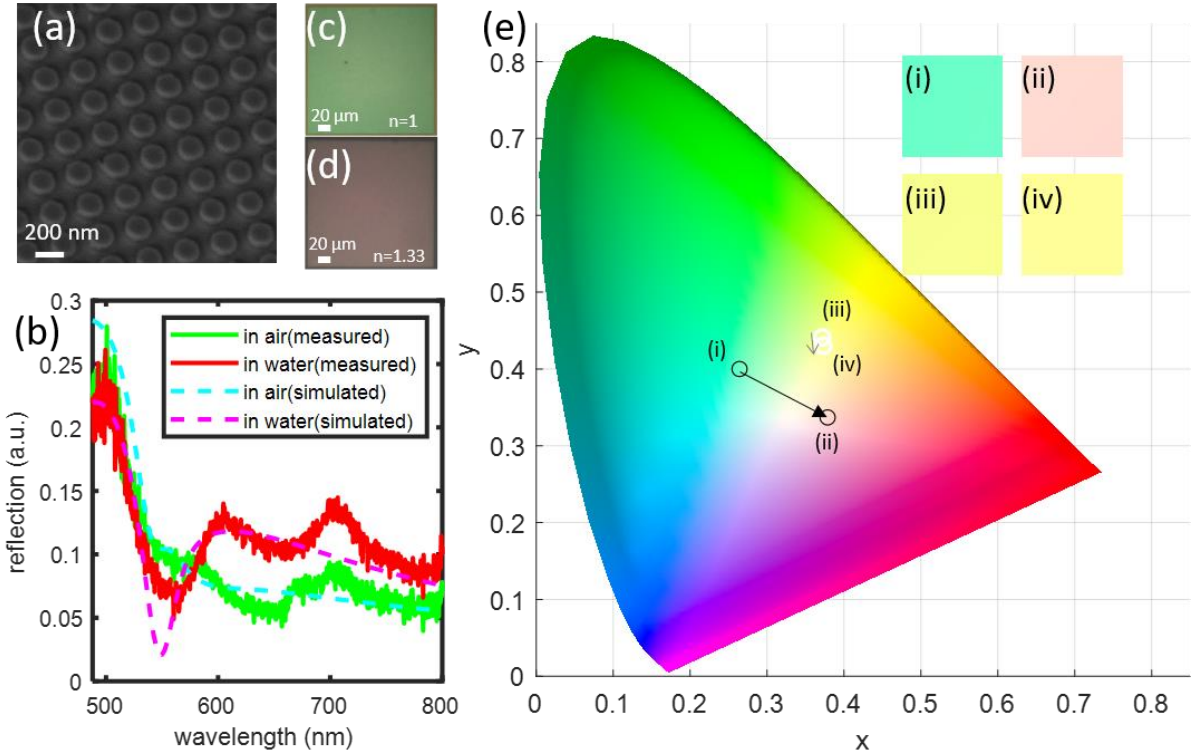

Figure 3: (a) SEM micrograph of the fabricated metasurface. (b) Measured and simulated reflection spectra of the metasurface. They show great agreement and indicate a vivid color change as expected when the environmental refractive index increases. (c) Optical images captured by digital camera when the metasurface is in air and (d) in water, respectively. The colors are consistent with the color predicted by measured spectra. (e) CIE 1931 color space plot showing the predicted color of the well-tailored color metasurface versus a random metasurface as a control group. The color metasurface shows green (i) in air and pink (ii) in water, while the control group shows only a minor chromatic shift, from (iii) to (iv). The four insets show the colors corresponding to the coordinates in color space labelled as (i) to (iv).

Once the color metasurface has demonstrated its ability to respond sensitively to the changes in the environmental refractive index, we proceeded to test the ability of this color metasurface to be integrated with optofluidics. Specifically, we focused on ETP flow in this work. After depositing chromium and gold thin layers for the generation of ETP flows, we reconfirmed the color of the metasurface using a digital camera to ensure that it maintains its color-changing abilities as the index is tuned, as shown in Fig. S3a and b.

The color metasurface was then integrated into a microfluidic channel (120 μm in height), and a 2-mW focused laser beam (973 nm) with a focus spot size of 1.33 μm was illuminated onto the metasurface to induce the heating effects. The calculated temperature rise is 10 K, making the final temperature in the microfluidic channel close to human body temperature and thus ensures the safety of delicate biological particles, such as proteins or EVs.

We then measured the in-plane ETP flow velocities. An a.c. electric field with a peak-to-peak voltage of 10 V (i.e., an electric field intensity of 83,333 V/m) and a frequency of 3 kHz was applied to generate the

ETP flow. As illustrated in Fig. 4b, the measured velocity of ETP flows can reach over 30 µm/s across a radial distance of 150 µm, showing good agreement with the simulation results in Fig. S1c. The ETP flow forms a micro-vortex[32] in the microfluidic channel to rapidly transport the suspended particles towards the hotspot. This active method of transporting particles overcomes the diffusion limits of Brownian motion and speeds up the incubation process as shown in the following. A more detailed Multiphysics modeling of the ETP flow is provided in Supplementary Information Section 1.

We then performed an experimental demonstration of detecting EVs in low concentration solution ($10^7$ EVs/mL). This concentration is below the normally reported concentration of EVs in blood, which is about $10^9$ to $10^{11}$ EVs/mL[33–37]. To easily capture the concentrated EVs, we functionalized the color metasurface with cysteamine molecules before injecting the EV solutions. Cysteamine is a widely employed agent for capturing EVs on gold surfaces[38,39] due to its positive charge. Additional details regarding the functionalization process can be found in the Supplementary Information.

The ETP flow was maintained for a 2-minute incubation by illuminating the metasurface with laser and applying an a.c. field of 3 kHz. The induced ETP flow transported EVs towards the illuminated metasurface where the negatively-charged EVs are captured on the cysteamine-functionalized metasurface. Subsequently, the moisture (i.e. water) was removed leaving behind the captured EVs and the sample imaged using a digital camera on an optical microscope. As depicted in Fig. 4c, a clear color change was perceived under the microscope at the position that had been illuminated by laser, suggesting the presence of EV clusters. The SEM image shown in Fig. 4d verifies the presence of EVs by showing nanoparticles with diverse sizes corresponding to the size range of EVs[40,41]. Fig. S3c depicts the color image of a metasurface region that has not been illuminated by the laser, where no color change occurs.

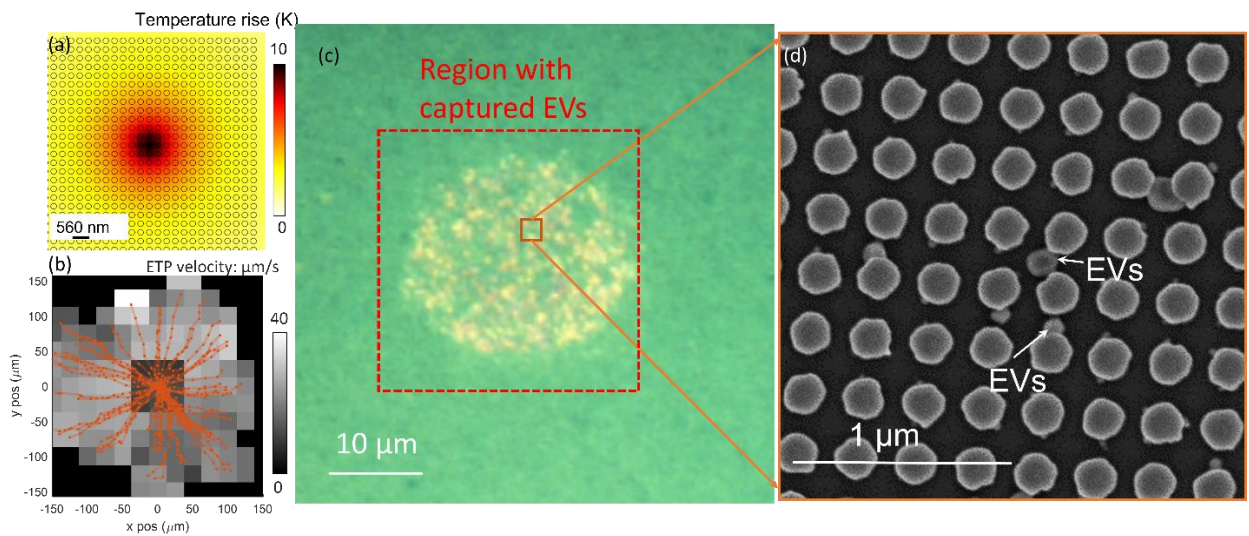

Figure 4: (a) Simulated temperature rise when using a 973 nm laser to illuminate on 3 nm Cr and 8 nm thick gold films coating on the color metasurface. The laser power is 2 mW and laser spot diameter is 1.33 µm, which are the values used in the experiments. (b) Trajectory map of EVs transported by the ETP flow with 2 mW laser illumination and 10 V a.c. electric field. The frequency of the a.c. electric field is 3 kHz. The background colormap shows the radial velocity magnitude of the ETP flow in plane, and the orange lines are the extracted particle trajectory. (c) Color image for the position where the laser was illuminated. It clearly shows local color changed due to the aggregation of EVs. (d) Zoomed-in SEM image for the region shown in (c). The SEM image clearly shows the existence of EVs captured on the optical metasurface.

**CONCLUSION:**

To summarize, in this work, we have successfully demonstrated an innovative colorimetric sensing approach based on silicon metasurface that yields structural coloration. This silicon metasurface exploits the response of multipolar optical modes to the changes in the refractive index of the surrounding to induce a vivid color change, allowing spectrometer-free and label-free sensing of extracellular vesicles. Furthermore, the combination of optofluidics and this color metasurface enables the detection of extracellular vesicles at femtomolar concentrations within a 2-minute incubation period.

As a proof-of-concept demonstration of the fusion of optofluidics and metasurface-based colorimetric sensing, we have showcased the capability of optofluidics to overcome the diffusion limit and established a platform with simple optical configurations. Notably, the aggregation of EVs by optofluidics driven by ETP flow in our configuration is highly efficient, and the EV assembly on the metasurface is easily perceivable to the human eye. One potential for future work to further push the limit of this technology for even lower concentration is to use computer vision or machine learning[12,42] for color image post-processing. While the initial demonstration employed the capture of EVs based on their surface charge through the use of cysteamine, the use of immuno-specific antibodieswould enable to capture and quantify disease-associated EVs by targeting disease-associated EV surface-bound protein markers and is a focus of future studies to build on this initial proof-of-concept.

**METHODS:**

EV solution preparation:

Unlabeled purified exosomes from COLO1 cell culture supernatant are purchased from Creative Diagnostics (DAGA-989) in a lyophilized phase. The lyophilized exosomes were dissolved into DI water and diluted to the desired concentration.

**ASSOCIATED CONTENT**

AUTHOR INFORMATION

**Corresponding Author**

*Justus C. Ndukaife

 justus.ndukaife@vanderbilt.edu

**Author Contributions**

JCN conceived the idea for this paper. CH and IH fabricated the sample. CH performed the experiments and conducted the numerical simulations. CH and SY performed the ETP flow analysis. CH and JCN composed the manuscript. JCN supervised the project.

**Acknowledgement:**

The authors acknowledge funding support from NSF CAREER Award (NSF ECCS 2143836).